\documentclass{article}
\usepackage{spconf,amsmath, amsfonts, graphicx,hyperref}

\title{PRECODER DESIGN IN MULTI-USER FDD SYSTEMS WITH VQ-VAE AND GNN}
\name{Srikar~Allaparapu, Michael~Baur, Benedikt~Böck, Michael~Joham, Wolfgang~Utschick \thanks{\copyright{20XX IEEE. Personal use of this material is permitted. Permission from IEEE must be obtained for all other uses, in any current or future media, including reprinting/republishing this material for advertising or promotional purposes, creating new collective works, for resale or redistribution to servers or lists, or reuse of any copyrighted component of this work in other works.}}}
\address{TUM School of Computation, Information and Technology, Technical University of Munich, Germany\\ Email: \{srikar.allaparapu, mi.baur, benedikt.boeck, joham, utschick\}@tum.de }

\begin{document}
\maketitle
\begin{abstract}
Robust precoding is efficiently feasible in frequency division duplex (FDD) systems by incorporating the learnt statistics of the propagation environment through a generative model. We build on previous work that successfully designed site-specific precoders based on a combination of Gaussian mixture models (GMMs) and graph neural networks (GNNs). In this paper, by utilizing a vector quantized-variational autoencoder (VQ-VAE), we circumvent one of the key drawbacks of GMMs, i.e., the number of GMM components scales exponentially to the feedback bits. In addition, the deep learning architecture of the VQ-VAE allows us to jointly train the GNN together with VQ-VAE along with pilot optimization forming an end-to-end (E2E) model, resulting in considerable performance gains in sum rate for multi-user wireless systems. Simulations demonstrate the superiority of the proposed frameworks over the conventional methods involving the sub-discrete Fourier transform (DFT) pilot matrix and iterative precoder algorithms enabling the deployment of systems characterized by fewer pilots or feedback bits.

\end{abstract}
\begin{keywords}
VQ-VAE feedback, precoder design, graph neural network, pilot optimization, measurement data
\end{keywords}
\section{Introduction} 
\label{sec:intro}

In the future generation of cellular communication systems (6G), a variety of technological innovations including massive multiple-input multiple-output (MIMO), intelligent reflective surfaces will be employed and artificial intelligence (AI) will play a vital role in enhancing the performances of these technologies by using data-driven, trained systems, resulting in positive consequences for energy efficiency and sustainability \cite{8869705}. The use of trained machine learning (ML) models for signal processing tasks such as channel estimation, equalization, and precoding enables further optimization compared to current 4G and 5G networks \cite{DLRole}, \cite{DLChEst}, \cite{VAEChEstMiBaur}.

In FDD systems, the uplink (UL) and downlink (DL) channel state information (CSI) is separated by a frequency gap. Thus, channel reciprocity cannot be assumed in general. A major problem in this context is the heavy CSI feedback overhead that arises after estimation of the DL CSI at the mobile terminal (MT), since precoding strategies usually assume the CSI knowledge at the base station (BS).
Predefined codebooks that are known at both the BS and MT are one approach to engage
this problem \cite{codebook1}, \cite{codebook2}. Such codebook-based approaches usually exhibit a low complexity nature, but are not site-specific, resulting in performance loss compared to data-based methods.

In this regard, many ML models have been proposed for CSI feedback estimation and compression \cite{DLCSIFeedback}, \cite{Vriz2}. One such novel approach is the vector quantized-variational autoencoder (VQ-VAE) \cite{nturanvqvae}. The VQ-VAE is a unique variant of a standard VAE that learns discrete representations in its latent space \cite{NIPS2017_7a98af17}. The idea is to fit a VQ-VAE to CSI data at the BS, share the encoder of the fitted VQ-VAE with the MT, and infer CSI feedback information by mapping it to the quantized VQ-VAE latent space. A significant advantage of the VQ-VAE-based strategy is that CSI estimation can be skipped and quantized feedback is generated depending solely on a noisy pilot observation.

On the other hand, utilizing the CSI feedback received at the BS, precoders are designed such that the total sum rate is maximized under a transmit power constraint. GNNs have been used for precoder design \cite{MDGNN}, \cite{CNNGNNPrecoding} replacing the iterative weighted minimum mean square error (WMMSE) algorithm \cite{IWMMSE}. Recently, a more practical approach, i.e., precoder design based on perfect statistical CSI \cite{StCSIICNet} and approximate statistical CSI using Gaussian
mixture models (GMMs) \cite{SrikarNTuran} have been found effective with the help of GNNs over the computationally intensive stochastic IWMMSE (SWMMSE) algorithms \cite{Razaviyayn}.

In this work, we replace the GMM in \cite{SrikarNTuran} by the VQ-VAE \cite{nturanvqvae} serving as an alternative generative prior. Further, with this incorporation of the VQ-VAE, a fully flexible E2E scheme which learns the CSI feedback module, and the precoder design module from the noisy input data has been investigated and realized. System setups with fewer pilots than transmit antennas are considered, rendering the instantaneous reconstruction of the CSI particularly challenging. Thus both, instantaneous reconstruction and statistical reconstruction as form of quantized CSI feedback have been investigated. The contributions of this paper can be summarized as follows :
\begin{itemize}
     \item We propose an E2E joint learning scheme of the VQ-VAE, the GNN, and the pilot matrix that together performs CSI quantization, robust precoder design, and pilot optimization with reduced training complexity.

     \item We explore both instantaneous as well as statistical CSI feedback reconstructions and examine the performance differences of one over the other. By leveraging model-based insights, we achieve a compact representation of the statistical information, enabling lightweight and efficient architectures.

    \item Simulations with real-world measurement data, producing statistical reconstructions, show the proposed frameworks' superiority over the methods from the referenced literature, including iterative algorithms and DFT codebook-based techniques, especially in systems with low pilot overhead.
\end{itemize}

\section{SYSTEM MODEL AND CHANNEL DATA}
\label{sec:systemmodel}

\subsection{Data Transmission} \label{sec:Data_Transmission}

We consider the DL of a single-cell multi-user system, where the BS is equipped with $N$ antennas and serves $J$ single-antenna MTs. The channel between MT $j$ and the BS is denoted by
$\boldsymbol{h}_j \in \mathbb{C}^N$. Linear precoding is applied and all users' precoders $\{\boldsymbol{v}_j\}_{j=1}^J$ are subjected to the transmit power constraint $\sum_j \|\boldsymbol{v}_j\|^2_2\leq \rho$. The instantaneous achievable sum rate to be maximized is given as
\begin{equation} \label{eq:sumrate}
    R = \sum_{j=1}^{J} \text{log}_2 \Biggl(1 + \frac{|\boldsymbol{h}_j^\mathrm{T} \boldsymbol{v}_j|^2}{\sum_{m \neq j} |\boldsymbol{h}_j^\mathrm{T} \boldsymbol{v}_m|^2 + \sigma_n^2} \Biggr)
\end{equation}
where $\sigma_n^2$ denotes the noise variance. The BS designs the precoders $\boldsymbol{v}_j$ based on each MT’s feedback information, which is encoded by $B$ bits.

\subsection{Pilot Transmission}
During the DL probing phase and before data transmission phase explained in Sec. \ref{sec:Data_Transmission}, MT $j$ receives $n_p$ channel observations, which are collected in 
\begin{equation} \label{eq:pilotobservation}
    \boldsymbol{y}_j = \boldsymbol{P} \boldsymbol{h}_j + \boldsymbol{n}_j
\end{equation}
where $\boldsymbol{n}_j \sim \mathcal{N}_\mathbb{C} (\boldsymbol{0}, \sigma_n^2 \operatorname{\mathbf{I}}_{n_p})$ represents the additive white Gaussian noise (AWGN). We either learn the pilot matrix or, unless otherwise mentioned, use a 2D-DFT (sub)matrix as the pilot matrix $\boldsymbol{P} \in \mathbb{C}^{n_p \times N}$ with unit norm columns, since in our simulations we consider a uniform rectangular array (URA) at the BS, cf.~\cite{DFTpilots}. We examine scenarios with fewer pilots than transmit antennas, i.e., $n_p < N$.

\subsection{Channel Data} \label{subsec:channeldata}
The measurements collected from a campaign conducted at the Nokia campus in Stuttgart, Germany, in 2017 are used in this work. The BS antenna with a down-tilt of 10\textdegree \ was installed on a rooftop approximately 20 m above the ground. It comprises a URA with $N_v = 4$ vertical ($\lambda$ spaced) and $N_h = 16$ horizontal antennas ($\lambda/2$ spaced), yielding in total $N = 64$ antenna elements. Here, $\lambda$ denotes the wavelength corresponding to the center frequency. The single monopole receive antenna, emulating MTs, was attached atop a mobile vehicle at a height of 1.5~m moving with maximum speed of 25~kmph. The carrier frequency employed was 2.18~GHz. For details, please refer to \cite{MeasurementCampaign}.

\section{PROPOSED FRAMEWORK}

\begin{figure*}[t]
\centerline{\includegraphics[width=\textwidth , height=6cm, keepaspectratio]{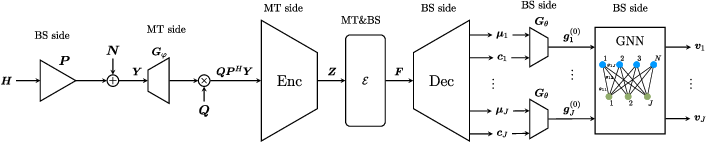}}
\caption{Structure of the proposed E2E model with learnable pilot matrix layer $\boldsymbol{P}$, VQ-VAE feedback and GNN precoding.}
\label{fig:e2emodel}
\end{figure*}

Fig.~\ref{fig:e2emodel} illustrates the proposed E2E transceiver model, distributed across the BS and the MTs, that jointly learns the pilot matrix, followed by the VQ-VAE network for the feedback scheme and then the GNN network of precoding. To simplify the design, the vectors for the different MTs $j=1,..,J$ are concatenated into matrices, as follows:
$\boldsymbol{H}=\{ \boldsymbol{h}_1, .., \boldsymbol{h}_J \}, \ \boldsymbol{N}=\{ \boldsymbol{n}_1, .., \boldsymbol{n}_J \}, \ \boldsymbol{Y}=\{ \boldsymbol{y}_1, .., \boldsymbol{y}_J \}, \ \boldsymbol{Z}=\{ \boldsymbol{z}_1, .., \boldsymbol{z}_J \}, \ \boldsymbol{F}=\{ \boldsymbol{f}_1, .., \boldsymbol{f}_J \}$. 

The input channels from the training dataset are first multiplied by the pilot matrix $\boldsymbol{P}$, which is modeled as a complex fully connected layer without bias, as used in \cite{ScalablePrecPilotOpt}. AWGN noise is added to form pilot observations as $\boldsymbol{Y} = \boldsymbol{P}\boldsymbol{H} + \boldsymbol{N}$. These observations are passed through a feature extractor $\boldsymbol{G}_\varphi$ at the MT side that acts as a coarse estimator for each MT $j$, the outputs of which are then multiplied by $\boldsymbol{Q}$, discussed in Sec.~\ref{sec:Structural_Insights}, to form the preprocessed input to VQ-VAE, see \cite{nturanvqvae}. The layout of the employed VQ-VAE in our work closely resembles the one in \cite{nturanvqvae}. The encoders at MTs generate latents $\boldsymbol{Z}$, which are quantized through embedding $\mathcal{E}$ shared between the MTs and BS, to feedback $\boldsymbol{F}$ according to the vector quantization operation described in Sec.~\ref{sec:Vector_Quantization}. After feeding back $\boldsymbol{F}$ to the BS, the BS generates MT-specific channel statistics $(\boldsymbol{\mu}_j, \boldsymbol{c}_j) \ \forall j$ by means of the decoder. These statistics can be then used for precoder design by the GNN network from \cite{SrikarNTuran} at the BS to obtain the precoders $\{\boldsymbol{v}_j\}_{j=1}^J$. The generated channel statistics $(\boldsymbol{\mu}_j, \boldsymbol{c}_j)$ from each MT $j$ are passed through a feature extractor $\boldsymbol{G}_\theta$ to obtain the permutation-equivariant features $\boldsymbol{g}_j^{(0)}$ required for the GNN processing. For further details on the VQ-VAE and GNN architectures, we refer to \cite{nturanvqvae}, \cite{SrikarNTuran}.

\subsection{Vector Quantization} \label{sec:Vector_Quantization}

Embedding $\mathcal{E}$ consists of $C$ elements, i.e., $\mathcal{E} = \{ \boldsymbol{e}_1, \ldots, \boldsymbol{e}_C\}$. Let the dimension of the codewords $\boldsymbol{e}_c \in \mathcal{E}$ be $N_E$, and let the latent space dimension be $N_L$. Then, the unquantized latent representation $\boldsymbol{z}_j$ is divided into $\frac{N_L}{N_E}$ sub-vectors. The $i$-th sub-vector $\boldsymbol{f}_{j,i}$ of the feedback vector $\boldsymbol{f}_j$ is obtained as
\begin{equation}
    \boldsymbol{f}_{j,i} = \underset{\boldsymbol{e}_c \in \mathcal{E}}{\text{arg min}}   \ ||\boldsymbol{e}_c - \boldsymbol{z}_{j,i} ||_2
\end{equation}
where $\boldsymbol{z}_{j,i}$ denotes the $i$-th sub-vector of $\boldsymbol{z}_j$, and $i = 1,..,\frac{N_L}{N_E} $. In this way,
the feedback vector $\boldsymbol{f}_j$ can be fully described by $B=\frac{N_L}{N_E}\text{log}_2C$ bits.

\subsection{Structural Insights} \label{sec:Structural_Insights}

Since the BS is equipped with a URA of $N=N_vN_h$ antenna elements, we confine the channel covariance matrices to be block-Toeplitz with Toeplitz blocks, which are expressed as $\boldsymbol{C}_j = \boldsymbol{Q}^\mathrm{H} \text{diag}(\boldsymbol{c}_j)\boldsymbol{Q}$, with $\boldsymbol{Q} = \boldsymbol{F}_{N_v} \otimes \boldsymbol{F}_{N_h}$ , where $\boldsymbol{F}_T$ (with $T \in \{N_v , N_h \}$) contains the first $T$ columns of a $2T \times 2T$ DFT matrix and $\boldsymbol{c}_j \in \mathbb{R}_{+}^{4N}$, cf. \cite{MeasurementCampaign2}, \cite{Statistical_Characterization}. Henceforth, $\boldsymbol{c}_j$ can be used to fully characterize $\boldsymbol{C}_j$ with an overall reduced dimension, which also then reduces the model complexity. 

\subsection{Training Scheme}

The training of the framework is done in a two-fold process, namely, the \textbf{\textit{pre-training stage}} and the \textbf{\textit{fine-tuning stage}}. Firstly, the training dataset is divided into two equal halves, one half is used in the pre-training stage and the other half in the fine-tuning stage, thereby reducing the training complexity of the framework. In the pre-training stage, the VQ-VAE network along with the pilot matrix is trained with the first half of the training dataset and this model is saved. Then in the fine-tuning stage, the saved VQ-VAE model along with the pilot layer is loaded and combined with the rest of the GNN network to be trained, \underline{\textit{without freezing}} the saved weights. The entire network is now trained based on the second half of the training dataset with a learning rate lower than the one used in pre-training. In this work, the fine-tuning learning rate is chosen as $\eta_{\text{fine-tune}} = \eta_{\text{pre-train}}/10$. The loss function used in the pre-training is the VQ-VAE loss from~\cite{nturanvqvae}, i.e., $\mathcal{L}_{\text{pre-train}} = \mathcal{L}_{\text{VQ-VAE}}$, given by
\begin{equation} \label{Eq:VQVAE_loss}
    \mathcal{L}_{\text{VQ-VAE}} = \mathcal{L}_{\text{rec}} + \|\text{sg}(\boldsymbol{z}_j) - \boldsymbol{f}_j\|_2^2 + \beta\|\boldsymbol{z}_j - \text{sg}(\boldsymbol{f}_j)\|_2^2
\end{equation} 
and since we model the conditional distributions of the VQ-VAE decoder to be Gaussian, the reconstruction loss is 
\begin{equation}
    \mathcal{L}_{\text{rec}} = \text{log det } (\pi \boldsymbol{C}_j) + (\boldsymbol{h}_j - \boldsymbol{\mu}_j)^\mathrm{H} \boldsymbol{C}_j^{-1} (\boldsymbol{h}_j - \boldsymbol{\mu}_j)
\end{equation}
whereas for fine-tuning, we chose the loss function to be the sum of VQ-VAE loss and the GNN loss which is the negative sum rate (NSR),  i.e., $\mathcal{L}_{\text{fine-tune}} = \mathcal{L}_{\text{VQ-VAE}} + \text{NSR}$.

As a benchmark, we also utilize the feedback scheme generating an instantaneous reconstruction $\boldsymbol{\bar{h}}_j$, cf. \cite{nturanvqvae}, utilizing the same VQ-VAE architecture but with decoder output covariance set to identity and only infer the mean vector as $\boldsymbol{\bar{h}}_j$. This variant is termed as VQ-AE in this work. In this case, the reconstruction loss is the MSE loss given by

\begin{equation}
    \mathcal{L}_{\text{rec}} = \|\boldsymbol{h}_j - \overline{\boldsymbol{h}}_j \|_2^2
\end{equation}

Due to the inherent capabilities of the VQ-VAE and GNN to generalize over different configurations with varying $J$ and SNR levels without requiring retraining \cite{SrikarNTuran}, \cite{nturanvqvae}, henceforth, this E2E training framework results in a single model containing learnt pilot matrix, infers the CSI feedback in quantized form, and generates the precoders.

\section{SIMULATION RESULTS}

We use 480,000 training data samples from the measurement campaign described in Sec. \ref{subsec:channeldata}. The data samples are normalized such that $\mathrm{E}[\|\boldsymbol{h}\|^2]=N=64$. To evaluate the performance, we use 10,000 testing data samples and the sum-rate metric is averaged over 500 multi-user constellations with $J$ MTs randomly selected from the evaluation set for each constellation. Additionally, we set $\rho =1$  in the precoding power constraint (cf. Sec. \ref{sec:Data_Transmission}) to enable the definition of SNR as $\frac{1}{\sigma_n^2}$. For the SWMMSE and the iterative WMMSE in the baselines, the maximum number of iterations is set to 300. We fix the VQ-VAE hyperparameters to $N_E=2, N_L=8, B=40 \ (C=1024)$, unless otherwise specified.

In the legends used, ``VQ-VAE(S) + GNN, learnt $\boldsymbol{P}$" and ``VQ-AE(I) + GNN, learnt $\boldsymbol{P}$" refer to the proposed E2E models with statistical inferences and instantaneous reconstructions respectively. ``VQ-VAE(S) + GNN" and ``VQ-AE(I) + GNN" refer to the baselines considered in this work, which represent the joint training of VQ-VAE/VQ-AE together with GNN without the pilot matrix learning. Instead here, the standard 2D-DFT (sub)matrix is set as $\boldsymbol{P}$. Furthermore as benchmarks, we use the schemes from \cite{nturanvqvae} using the VQ-VAE feedback scheme with precoding algorithms denoted by ``VQ-VAE(S) + SWMMSE" and ``VQ-AE(I) + WMMSE" as well as ``GMM + GNN" and ``GMM + SWMMSE" from \cite{SrikarNTuran}.

\begin{figure}[t]
\centerline{\includegraphics[width=\columnwidth , height=6cm, keepaspectratio]{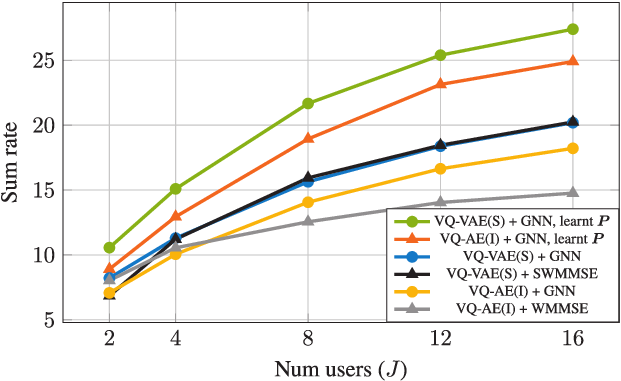}}
\caption{Sum rate over the number of MTs $J$ for a system with $\text{SNR}=15$dB, $B=40$ bits and $n_p=8$ pilots }
\label{fig:sr_vs_J}
\end{figure}

In Fig. \ref{fig:sr_vs_J}, we fix $\text{SNR}=15 \ $dB, $B=40$ bits, $n_p=8$ pilots and vary the number of MTs $J$. Note that $N=64$. We observe a significant boost in performances by the E2E models in both the cases of statistical and instantaneous reconstructions over all the baseline schemes, caused by jointly learning the pilot matrix replacing the standard sub-DFT matrix, which is not scenario-specific, together with the VQ-VAE feedback scheme and GNN-based precoder design, benefitting from the advantage of huge complexity reduction over conventional schemes and algorithms, which are rather computationally intensive. Furthermore, we observe that the E2E scheme of ``VQ-VAE(S) + GNN, learnt $\boldsymbol{P}$" performs the best, indicating the importance of deploying statistical inference-based learning models and frameworks, particularly in systems with low pilot overhead.

\begin{figure}[t]
\centerline{\includegraphics[width=\columnwidth , height=6cm, keepaspectratio]{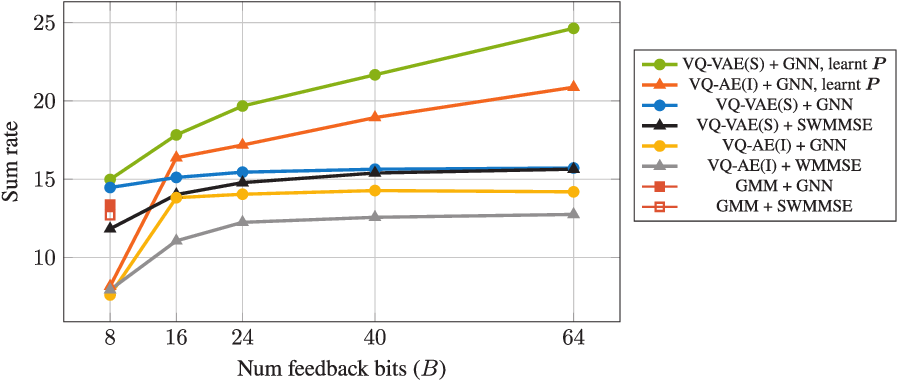}}
\caption{Sum rate over the number of feedback bits $B$ for a system with $\text{SNR}=15$dB, $J=8$ MTs and $n_p=8$ pilots}
\label{fig:sr_vs_B}
\end{figure}

In Fig.~\ref{fig:sr_vs_B}, we examine the effect of the number of feedback bits $B$ on the sum rates with fixed $\text{SNR}=15 \ $dB, $J=8$ MTs and $n_p=8$ pilots. The proposed ``VQ-VAE(S) + GNN, learnt $\boldsymbol{P}$" framework performs the best with increasing $B$. Here we also observe that with increasing $B$, the sum rates of ``VQ-VAE(S) + GNN, learnt $\boldsymbol{P}$" and ``VQ-AE(I) + GNN, learnt $\boldsymbol{P}$" steadily increase in contrast to the corresponding counterparts employing sub-DFT $\boldsymbol{P}$ where they exhibit saturation beyond $B\geq 24$ bits. The E2E schemes benefit from the additional feedback knowledge with increased yet practical codebook sizes at higher $B$ to design site-specific pilots accordingly, which is not the case for the standard pre-defined sub-DFT pilot matrix. Here, we also present the performance of the GMM baseline only for $B=8$ bits as the GMM model suffers from overfitting issues for $B > 9$ bits. The VQ-VAE scheme combined with GNN outperforms the GMM scheme with GNN as well as SWMMSE proving the effectiveness of the proposed E2E model.

\begin{figure}[t]
\centerline{\includegraphics[width=\columnwidth , height=6cm, keepaspectratio]{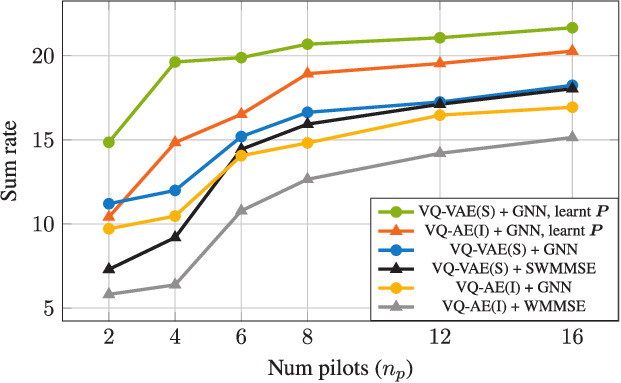}}
\caption{Sum rate over the number of pilots $n_p$ for a system with $\text{SNR}=15$dB, $J=8$ MTs and $B=40$ bits}
\label{fig:sr_vs_np}
\end{figure}

In Fig. \ref{fig:sr_vs_np}, we assess the impact of the number of pilots on the system performances by setting $\text{SNR}=15 \ $dB, $J=8$ MTs and $B=40$ bits. We observe that the proposed E2E model of ``VQ-VAE(S) + GNN, learnt $\boldsymbol{P}$" performs the best for all considered numbers of pilots $n_p$. With only $n_p=4$ pilots, it achieves the sum rate that ``VQ-AE(I) + GNN, learnt $\boldsymbol{P}$" attains at $n_p=16$ pilots. This analysis further accentuates that the E2E learning-based systems with lower pilot overhead can be deployed without sacrificing performance compared to the baseline schemes by means of statistical modeling.

\section{CONCLUSION AND OUTLOOK}
\vspace{-0.3cm}
This work presents an E2E learning based transceiver framework that jointly learns the pilot matrix, designs feedback and precoders, modeled based on statistical inference. We also proposed a novel training scheme by dividing the training into pre-training and fine-tuning stages, with reduced training complexity by using only half of the training dataset in each stage. By leveraging model-based structural insights, we make use of a compact representation of the statistical knowledge that results in reduced model complexity and enhanced system performance, especially in systems with low pilot overhead. As a potential future work, we suggest to take up a redesign of the VQ-VAE-based feedback scheme and refined codebook learning procedure which is adaptable to dynamic feedback capacities, i.e., generalizes over $B$.


\bibliographystyle{IEEEbib}

\end{document}